# European research on magnetic nanoparticles for biomedical applications: standardisation aspects

Peter Schier[1], Craig Barton[2], Simo Spassov[3], Christer Johansson[4], Daniel Baumgarten[1], Olga Kazakova[2], Paul Southern[5], Quentin Pankhurst[5], Marco Coisson[6], Cordula Grüttner[7], Alex Price[8], Roman Rüttinger[9], Frank Wiekhorst[10], James Wells[10] and Uwe Steinhoff[10]

[1] UMIT - Private University for Health Sciences, Medical Informatics and Technology, 6060 Hall in Tirol, Austria
[2] National Physics Laboratory, Teddington TW11 0LW, United Kingdom
[3] Institut Royal Météorologique De Belgique, 1180 Bruxelles, Belgium
[4] RISE Research Institutes of Sweden AB, 400 14 Göteborg, Sweden
[5] University College London, London WC1E 6BT, United Kingdom
[6] Istituto Nazionale di Ricerca Metrologica, 10135 Torino, Italy
[7] micromod Partikeltechnologie GmbH, 18119, Rostock, Germany
[8] BSI Standards Limited, London W4 4AL, United Kingdom
[9] DIN Deutsches Institut für Normung e. V., 10787 Berlin, Germany
[10] Physikalisch-Technische Bundesanstalt, 10587 Berlin, Germany
`peter.schier@umit.at`

**Abstract.** Magnetic nanoparticles have many applications in biomedicine and other technical areas. Despite their huge economic impact, there are no standardised procedures available to measure their basic magnetic properties. The International Organization for Standardization is working on a series of documents on the definition of characteristics of magnetic nanomaterials. We review previous and ongoing European research projects on characteristics of magnetic nanoparticles and present results of an online survey among European researchers.

## 1 Introduction

Liquid suspensions of magnetic nanoparticles (MNPs) are used in many technical areas like loudspeakers, mobile phones, vacuum sealings, metal separation and water remediation. In the biomedical applications, MNPs play a very important role in in-vitro diagnostics for the separation of cells, bacteria, viruses, protein, nucleic acids and other compartments from blood and body liquids. MNPs are used as contrast agents in Magnetic Resonance Imaging, as well as tracers in Magnetic Particle Imaging and sentinel lymph node detection as well as in MNP based therapies, where they act as heating agents in magnetic field hyperthermia or as drug carriers in magnetic drug targeting and magnetic gene therapies [1]. The economic impact of MNP based biomedical products of European companies alone amounts so far to more than 2 billion € per year [2]. The largest part of this economic impact is generated by in-vitro diagnostics applications.

Another large application field is the use of nanostructured iron oxide in pigments for cosmetics, structural engineering, and many other purposes. The annual production of those pigments alone in the European Union is more than 100,000 tons per year [3]. Obviously, this creates a demand for international standards on the main characteristics of MNPs and the respective measurement procedures.

The International Organization for Standardization (ISO) currently prepares document standards on magnetic nanomaterials to provide harmonised definitions for commercial trade, application development, regulation and science in the MNP sector. Current activities focus on magnetic nanosuspensions (ISO 19807-1) [4] and magnetic beads for DNA extraction (ISO 19807-2) [5].

In a new field such as nanotechnology, the foremost interest is the definition of terms and characteristics. These terms can then be used by MNP manufacturers for labelling their products and designing technical data sheets, or by scientists exchanging information on MNPs. While the definitions for chemical composition and mechanical fluid properties could be taken from existing standard documents, many of the magnetic properties of MNP suspensions needed a new definition.

The selection of important MNP characteristics should be based on reviews of the scientific work on MNPs and of economically relevant MNP based applications. Outstanding in this field have been two pre-normative European projects: The EU FP7 project NanoMag, devoted to metrological groundwork on MNP standardisation and the EU COST Action RADIOMAG, that dealt with the standardisation of the measurement of MNP hyperthermia performance. Both projects will be briefly summarized in this paper. Furthermore, this paper is a result of the co-normative EURAMET project MagNaStand, which is supporting the ISO standardisation of MNP. In the framework of MagNaStand, we have evaluated the European CORDIS database for previous and running EU-projects concerned with MNP. In addition to the pure database work, we have also performed a survey of researchers involved with MNPs.

## 2 Pre- and co-normative EU projects on magnetic nanoparticles

### 2.1 The pre-normative EU NanoMag project

The NanoMag project was funded by the EU FP7 research program in the years 2013 – 2017 and had a budget of about 11 M€. The NanoMag project objectives were to standardise and harmonise ways to measure and analyse the data for MNP systems. NanoMag brought together leading experts in: synthesis of magnetic single- and multi-core nanoparticles, characterisation of magnetic nanoparticles, and national metrology institutes. The project defined standard measurements and techniques which are necessary for defining a magnetic nanostructure and quality control [6]. NanoMag was focused on biomedical applications, for instance bio-sensing (detection of different biomarkers), contrast substance in tomography methods (Magnetic Resonance Imaging and Magnetic Particle Imaging) and magnetic hyperthermia (for cancer therapy).

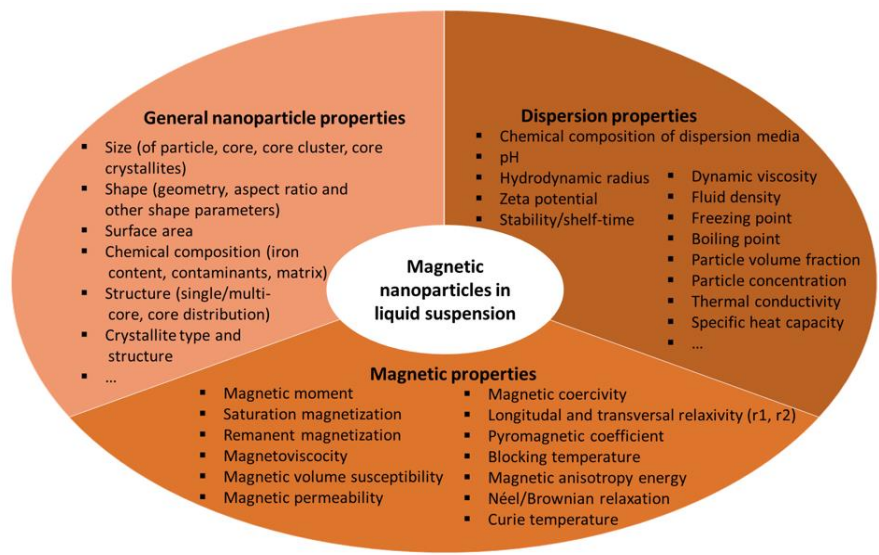

**Fig. 1.** Important parameters of liquid MNP suspensions that can be standardised.

Initially, NanoMag was designed as a pre-normative project for the preparation of measurement standards. When the International Organization for Standardization started a normative activity for material properties of MNP suspensions, NanoMag members entered the ISO committee TC 229 "Nanotechnologies" as technical experts and made substantial contributions to ongoing ISO standardization work in the field of MNP systems.

The knowledge of the NanoMag experts has been summarized in four electronic-learning modules that can be accessed via the internet [7].

To support MNP standardization the NanoMag project has:

- developed a clearer nomenclature to describe the structure and magnetic properties of MNPs and MNP ensembles (see Fig. 1);
- performed surveys of measurement methods for MNPs and their pros and cons, classification of these methods including also classification of different types of MNPs (single-core, multicore, nanoflowers);
- used Monte-Carlo simulations to explain the experimental results, which have led to improved modelling of MNP magnetic properties, especially the dynamic magnetic behaviour; and
- performed initial round robin measurements (using the same analysis methods on the same MNP samples but in different laboratories utilizing the developed standard operating procedures (SOPs)) using different types of MNP systems.

A standardization roadmap developed during the NanoMag project forecasted the development of measurement standards for MNP to start in the year 2021. Many results of the NanoMag project were summarized in a review paper in [8].

## 2.2    EU Transdomain COST Action RADIOMAG

The EU COST Action RADIOMAG was a network of over 140 scientists in the field of tumour therapy and it was dedicated to research on multifunctional nanoparticles for Magnetic Field Hyperthermia and Indirect Radiation Therapy. The RADIOMAG project duration was from 2014 – 2018. An important task of RADIOMAG was the harmonisation and standardisation of the assessment of heat generation by MNP in an alternating magnetic field, since this is the basic principle in Magnetic Field Hyperthermia [9].

Magnetic Field Hyperthermia, especially in combination with radiotherapy, has been demonstrated as an effective tool to slow down or stop tumour growth and to support anti-cancer therapy in difficult tumour cases like glioblastoma. Pilot studies in humans are underway in a number of clinics in Europe. In April 2019 a new magnetic field hyperthermia treatment centre for brain tumours has been opened at the Independent Public Clinical Hospital No. 4 (SPSK 4) in Lublin, Poland [10].

These continuous developments have also created a niche market for several small and medium-sized enterprises for manufacturing magnetic field hyperthermia test devices. These devices are mainly used by academics testing in-vitro and ex-vivo the efficiency of MNP suspensions to deliver heat. The relevant physical quantity is the Specific Loss Power (SLP). For calorimetric determinations, the SLP is deduced from time vs. temperature curves T(t), measured with the SLP test device.

The RADIOMAG activities concerning the standardized characterization of MNP for magnetic field hyperthermia focused on SLP measurements:

1. Survey amongst RADIOMAG members on available SLP test devices/setups and their field/frequency combinations;
2. Development of a standard operating procedure for the calorimetric SLP determination, i.e. from T(t) measurements;
3. A comparative SLP determination on water-based ferrofluids between 21 participating laboratories (SLP ring test) and evaluation of SLP calculation methods;
4. Study of the field dependence, i.e. SLP(H); and
5. Design of a possible calibration sample for SLP test devices.

The results showed that a large majority of groups determine the SLP from calorimetric measurements with non-adiabatic setups, also commercially available on the market. In contrast, only a few producers exist for non-calorimetric devices using AC hysteresis and a single laboratory used a "home-made" nearly adiabatic setup.

The RADIOMAG work demonstrated that there are no common procedures available for carrying out T(t) measurements in magnetic field hyperthermia setups. Typically, different laboratories use their own individual best-practice protocols, or follow instructions given in the SLP test device manual, in case of commercial setups. RADIOMAG has therefore developed a SOP for measurements, including a questionnaire for instrument specific parameters.

Furthermore, RADIOMAG performed a ring test on the determination of SLP values, where the same MNP formulation was investigated by 21 different laboratories. The results showed a significant variation of quantitative SLP values, even for an identical MNP suspension. Further analysis of these results is still ongoing, and the publication of detailed results is expected in 2019/2020.

### 2.3 The EURAMET project MagNaStand

The EURAMET organisation has established the co-normative project MagNaStand, running from 2017 – 2020. The objective of MagNaStand is to collect the available knowledge on standardised measurements of MNPs, to create it where it is not readily available, to make this knowledge available for the standardisation of MNPs at ISO level, and to involve stakeholders from industry and academia.

Specifically, this includes the preparation of standard operating procedures for static and dynamic magnetisation measurements and for specific loss power assessment in magnetic hyperthermia. The MagNaStand project works on the definition of long-term stability of MNP suspensions and of SLP in magnetic hyperthermia in a metrologically sound and traceable way.

Another objective of the MagNaStand project is the preparation of future measurement standards for MNP by summarizing the existing metrological knowledge in a structured form for static magnetic susceptibility, dynamic magnetic susceptibility and specific loss power in magnetic hyperthermia.

The MagNaStand projects enables the continuous participation of European experts in the standardisation work at ISO/TC229. Thus, the scientific results of the pre-normative EU projects "NanoMag" and "RADIOMAG" can be introduced into the international standardisation process. This includes the definition of terms for magnetic quantities and the compartments of magnetic nanoparticles, actual versions of SOPs for magnetic measurements, and surveys of industrial requirements on standardisation of magnetic nanoparticles. MagNaStand experts are co-leaders of the development of ISO/TS 19807-2 "Magnetic beads for DNA extraction", together with experts from China.

## 3 Search of the EU CORDIS database for research projects that are relevant for MNP standardization

In this chapter we present the evaluation of the Community Research and Development Information Service (CORDIS) database on research projects funded by the EU. The aim of this activity was to provide a short summary of the most important EU FP7 and H2020 projects on MNPs. The search terms are explained in Table 1.

### 3.1 Results of the search of the CORDIS database

The search of the CORDIS website was conducted using all search terms of Table 1 in quotation marks and connected by Boolean OR operators (i.e. "*magnetic *particle" OR "*magnetic *bead*" OR etc.).

Afterwards the result was filtered by the homepage's "Refine by:" option. Only results of the "Content Type: Project" which were part of "Programme: Horizon 2020" or "Programme: FP7" were downloaded in tabular form as CSV-files and - for the full description of the project objectives – as PDF-booklets. The search resulted in a total of 108 EU research projects with summarized project budgets of roughly € 267 million. In comparison, a search in the Research Portfolio Online Reporting Tools (RePORT) of the U.S. Department of Health & Human Services using the same search terms from Table 1 resulted in 214 projects with a total budget volume of $ 67 million. A close inspection of the search results revealed that despite the wide range of the search terms, a number of relevant EU projects concerned with MNPs that were already known to the authors were not captured. They were later added manually to the final list of relevant EU projects concerned with MNP. The final total number of EU projects concerned with MNP was 118 with a total budget of € 348 million.

**Table 1.** Search terms in the CORDIS database for identification of relevant EU projects for MNP standardisation.

| Search term | Description |
|---|---|
| *magnetic *particle* | Since the standardisation of magnetic nanoparticle characteristics is our main goal, it is only natural to find all projects containing any variant form of "*magnetic *particle*" (e.g. "superparamagnetic nanoparticles"). |
| *magnetic *bead* | Some research groups use the notation "bead" instead of "particle". Otherwise, the reason to use this search term is the same as "*magnetic *particle*". |
| iron oxide nanoparticle | Iron oxide nanoparticles are by far the most commonly used magnetic nanoparticles. |
| superparamagnetic* | A salient feature of MNPs is their superparamagnetic behaviour. |

## 4 An online survey of researchers involved in MNP projects

We compiled a list of contact addresses of the leaders of the identified EU MNP projects, enhanced by contacts from the NanoMag and RADIOMAG networks. Altogether, over 100 European researchers were asked to participate in an online survey on MNP standardisation. We have received 32 responses, of which the most relevant results are summarized in this chapter in a question-answer scheme.

### 4.1 Survey results

**Q. 01**: Which of the following were your application areas of MNP? **A. 01**: 32 biomedical applications, 14 MNP synthesis, 6 environmental applications, 9 pharmaceutical applications, 6 other.

**Q. 02**: What was the Technology Readiness Level (TRL) of your MNP project? **A. 02**: 22% TRL1 – basic principles observed and reported, 19% TRL2 – technology concept and/or application formulated, 30% TRL3 – characteristic proof-of-concept, 23% TRL4 – component validation in lab environment, 2% TRL5 – component validation in relevant environment, 2% TRL6 – prototype demonstration in relevant environment, 1% TRL7 – prototype demonstration in application environment, and 0 in TRL8 – complete system in test and demonstration and TRL 9 – complete system in successful operation.

**Q. 03**: What kinds of materials for MNP were used during the project? Please specify core- and coating material. A free-text answer was possible. **A. 03**: Magnetite (13) and maghemite (9) was the most common core material, other ferrites (Barium-Ferrite, Nickel-Ferrite, Nickel-Cobalt-Ferrite, etc.) (10) played also a role. The coating material came from these material groups:

- Polymer/Organic which included plastics, polysaccharides and organic acids (53)
- Metal/Alloy which included metals, metalloids and alloys (10)
- Biofunctionalized which included bacteria and proteins (7)
- No coating for uncoated MNPs (1)

**Q. 04**: In which environment(s) were the MNP used? **A. 04**: 43% laboratory, 34% in-vitro (cell cultures), 25% in-vivo.

**Q. 05**: Properties of the applied magnetic field: specify field strength and frequency during your MNP application. **A 05**.: see Fig. 2

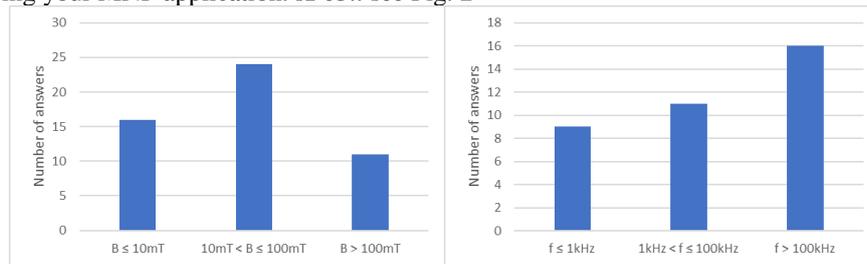

**Fig. 2.** Magnetic field strength and frequency during the MNP application.

**Q. 06**: Did you use a static magnetic field or an alternating magnetic field? **A.06**: 25 alternating magnetic field, 18 static magnetic field

**Q. 07**: What are the sources of the MNP characteristics that you used, please rank. **A. 07**: Summary rank: 1. own measurements, 2. literature values 3. technical data sheet, 4. custom measurement.

**Q. 08**: Please rate the importance of proposed characteristics for your project (1=low, 5=high). **A.08**: see Fig. 3

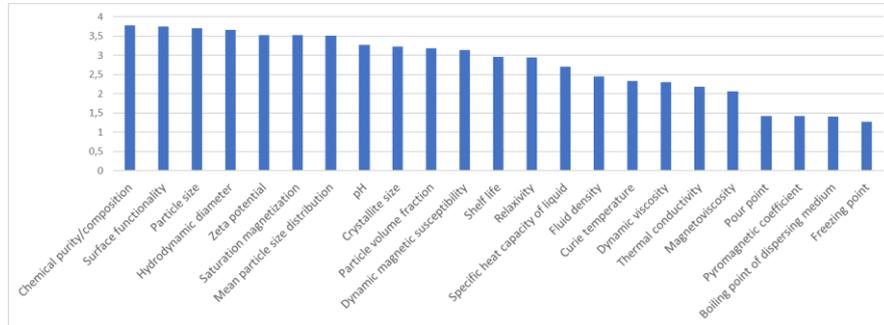

**Fig. 3.** Importance of different MNP parameters.

**Q. 09**: Please enter the approximate value range of the characteristics. **A. 09**: see Fig. 4

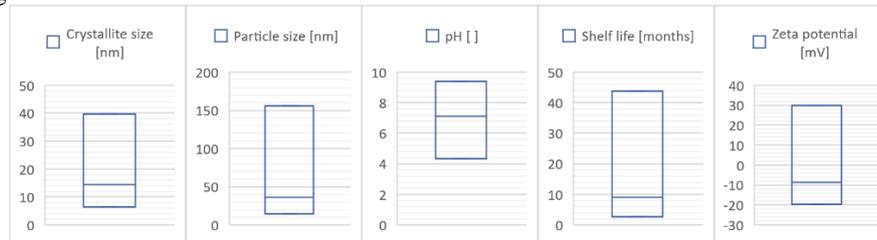

**Fig. 4.** Mean value ranges for several MNP parameters (minimum, typical and maximum).

**Q. 10**: Did you encounter any serious problems during the measurement process of important characteristic properties? **A. 10**: 64% no, 21% unclear measurement procedure, 15% measurements were not reproducible.

**Q. 11**: Are you aware of reference laboratories, where you could check or cross-validate your own measurement results? **A. 11**: 50% yes, 37% no, 13% no answer.

**Q. 12**: Please rank the most important measurement technique to characterize MNP. **A. 12**: 1. transmission electron microscopy (TEM), 2. magnetorelaxometry (MRX), 3. Hysteresis loops (DC magnetometry), 4. ZFC/FC curves (temperature dependent DC magnetometry) and 5. X-ray powder diffraction (XRD).

**Q. 13**: What are the most important characteristics? **A. 13**: 36 particle size, 27 saturation magnetisation, 13 hydrodynamic size, 9 specific absorption rate, 8 stability of suspension, 8 chemical composition, 6 biological properties, 5 dispersity, all other <5.

**Q. 14**: Did you use a standardized measurement protocol or were there any other standardisation aspects in your project? **A. 14**: 50% no, 25% yes, 25% no answer

**Q. 15**: Did you encounter any problems with the used MNP due to erroneous/unspecified characteristics? If yes, please elaborate. **A. 15**: People were dissatisfied with numerous things involving erroneous or unspecified characteristics of MNPs. Most of the complaints can be summarized by stating that the information provided by manufacturers was incomplete. Complaints involved missing expiration dates, wrong or missing concentration values, wrong or missing magnetite/maghemite indications, unknown surface compositions, stability change over time and recommended storage conditions. Several participants also described problems reproducing previously obtained or published results.

**Q. 16**: How would you rate the current state of standardization of MNP? (1=low, 4=high). **A. 16**: see Fig. 5.

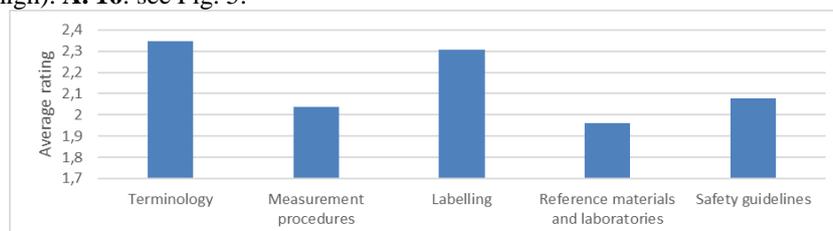

**Fig. 5.** Rating of current standardisation of different aspects of MNP handling.

### 4.2 Discussion of the survey results

The survey, although it is still based on a limited data base, shows clearly the wide range of applications and parameters that have to be considered when the standardisation of MNP is intended. It reveals that EU funded MNP research so far is restricted to very early or early Technology Readiness Levels. This might be one reason that the need for standardisation has been underestimated so far. On the other hand, the survey demonstrates that even in early research in scientific labs people often encounter problems with non-standardised parameters and procedures and non-reproducible measurement results. The need for further standardisation is clearly demonstrated. The survey gives valuable hints, which measurement methods and parameters are most important and would most benefit from standardised procedures.

## 5 Summary

Despite the huge economic importance of magnetic nanoparticles, there exists no standardised description and measurement of their basic magnetic properties. This is especially surprising with respect to the 118 research projects and the budget of € 348 million that the European Union has invested into MNP related projects since the FP7 program. The pre- and co-normative EU projects NanoMag, RADIOMAG and MagNaStand have made decisive steps to change this situation. Currently, Europe is an important contributor to the standardisation of magnetic nanomaterials that is performed in the ISO/TC229 "Nanotechnologies" committee. The efforts for harmonisation are based on surveys and other interactions with academic and industrial stakeholders.

We have presented the detailed content of such a survey in order to stimulate a discussion on the most urgent needs in MNP metrology. In addition to the basic description that is currently developed at ISO, standard operating procedures for magnetic measurements, calibrated measurement devices and certified reference materials for magnetic MNP properties are desirable outcomes of future research.

## 6     Acknowledgments

This work was supported by the EMPIR program co-financed by the Participating States and from the European Union's Horizon 2020 research and innovation program, grant no. 16NRM04 "MagNaStand".